\documentstyle[sprocl,epsfig]{article}

\input{psfig}

\def\Journal#1#2#3#4{{#1} {\bf #2}, #3 (#4)}

\def\NPB{{\em Nucl. Phys.} B}
\def\PLB{{\em Phys. Lett.}  B}
\def\PRL{\em Phys. Rev. Lett.}
\def\PRD{{\em Phys. Rev.} D}
\def\ZPC{{\em Z. Phys.} C}
\def\PREP{\em Phys. Rep.}

\def\be{\begin{equation}}
\def\ee{\end{equation}}
\def\bea{\begin{eqnarray}}
\def\eea{\end{eqnarray}}

\begin{document}

\title{THE CKMT MODEL AND THE THEORETICAL DESCRIPTION
OF THE CALDWELL-PLOT
\footnote{Contribution to the Proceedings
of the XVII Autumn School on QCD: Perturbative or Nonperturbative?,
29 September-4 October 1999, IST, Lisbon (Portugal), edited by 
L.S.~Ferreira, P. Nogueira, and J.I. Silva-Marcos, World Scientific.}}

\author{C. MERINO}

\address{Dpto. de F\'\i sica de Part\'\i culas,\\
Universidade de Santiago de Compostela,\\
E-15706 Santiago de Compostela, Spain}

\author{A.B. KAIDALOV}

\address{ITEP, B. Cheremushkinskaya 25,\\
117259 Moscow, Russia}

\author{D. PERTERMANN}

\address{Physics Department, Univ-GH-Siegen,\\
D-57068 Siegen, Germany}


\maketitle\abstracts{The CKMT model for the nucleon structure
function $F_2$ is
in good agreement with the HERA data at low and moderate $Q^2$. 
The fit to the same data obtained with a modified version of the 
model in which
a logarithmic dependence on $Q^2$ has been included is also discussed. 
Finally, we
show how the parametrization of the CKMT model for the nucleon 
structure function $F_2$ describes the HERA data when presented
in the Caldwell-plot as the behavior of the
logarithmic slopes of $F_2$ vs $x$ and $Q^2$.}

\section{The CKMT model}
The CKMT model \cite{ckmt} for the parametrization of the nucleon structure 
function
$F_2$ is a theoretical model based on Regge theory which provides a consistent
formulation of this function in the region of low $Q^2$, and can
be used as a safe and theoretically justified initial condition in the
perturbative QCD evolution equation to obtain the structure function at larger
values of~$Q^2$. 

The CKMT model \cite{ckmt} proposes
for the nucleon structure functions
\begin{equation}
F_2(x,Q^2) = F_S(x,Q^2) + F_{NS}(x,Q^2),
\label{eq:eq1}
\end{equation}
the following parametrization of its two terms in
the region of small and moderate $Q^2$. For the singlet term, corresponding to
the Pomeron contribution:
\begin{equation}
F_S(x,Q^2) = A\cdot x^{-\Delta(Q^2)}\cdot(1-x)^{n(Q^2)+4}
\cdot\left({Q^2\over Q^2+a}\right)^{1+\Delta(Q^2)},
\label{eq:eq2}
\end{equation}
where the $x$$\rightarrow$0
behavior is determined by an effective intercept 
of the Pomeron,~$\Delta$,
which takes into account Pomeron cuts and, therefore (and this is one of the
main points of the model), it depends on $Q^2$. This dependence was 
parametrized~\cite{ckmt}
as :
\begin{equation}
\Delta (Q^2) = \Delta_0\cdot\left(1+{\Delta_1\cdot Q^2
\over Q^2+\Delta_2}\right).
\label{eq:eq3}
\end{equation}
Thus, for low values of $Q^2$ (large cuts), $\Delta$ is close
to the effective value found from analysis of hadronic total cross-sections
($\Delta$$\sim$0.08), while for high values of $Q^2$ (small cuts),
$\Delta$ takes the bare Pomeron value,
$\Delta$$\sim$0.2-0.25. The
parametrization for the non-singlet term, which corresponds to the secondary
reggeon (f, $A_2$) contribution, is:
\begin{equation}
F_{NS}(x,Q^2) = B\cdot x^{1-\alpha_R}\cdot(1-x)^{n(Q^2)}
\cdot\left({Q^2\over Q^2+b}\right)^{\alpha_R},
\label{eq:eq4}
\end{equation}
where the $x$$\rightarrow$0 behavior is determined by the secondary
reggeon intercept $\alpha_R$, which is in the range $\alpha_R$=0.4-0.5.
The valence quark contribution can be separated into the contribution of the 
u, $B_u$, and d, $B_d$, valence quarks,
the normalization condition for valence quarks fixing 
both contributions 
at one
given value of $Q^2$ (we use $Q_0^2=2.GeV^2$ in our calculations).
For both the singlet and the non-singlet terms, the behavior when
$x$$\rightarrow$1 is
controlled by $n(Q^2)$, with $n(Q^2)$ being
\begin{equation}
n(Q^2) = {3\over2}\cdot\left(1+{ Q^2
\over Q^2+c}\right),
\label{eq:eq5}
\end{equation}
so that, for $Q^2$=0, the valence quark distributions have the 
same power, given by
Regge intercepts, as in Dual
Parton Model \cite{dpm}, $n$(0)=$\alpha_R$(0)$-$$\alpha_N$(0)$\sim$ 3/2, and 
the 
behavior of
$n(Q^2)$ for large $Q^2$ is taken to coincide with 
dimensional counting rules.

The total cross-section for real ($Q^2$=0) photons can be obtained from the
structure function $F_2$ using the following relation:
\begin{equation}
\sigma^{tot}_{\gamma p}(\nu) = \left[{4\pi^2\alpha_{EM}\over Q^2}
\cdot F_2(x,Q^2)\right]_{Q^2=0}.
\label{eq:eq6}
\end{equation}
The proper $F_2(x,Q^2)$$\sim$$Q^2$
behavior when
$Q^2$$\rightarrow$0, is fulfilled in the 
model
due to the last factors in equations \ref{eq:eq2} and \ref{eq:eq4}.
Thus, the
$\sigma^{tot}_{\gamma p}(\nu)$ has the following form in
 the CKMT model:
\begin{equation}
\sigma^{tot}_{\gamma p}(\nu) = 4\pi^2\alpha_{EM}
\cdot\left(A\cdot a^{-1-\Delta_0}\cdot(2m\nu)^{\Delta_0}
+(B_u+B_d)\cdot b^{-\alpha_R}\cdot(2m\nu)^{\alpha_R-1}\right).
\label{eq:eq6a}
\end{equation} 

The parameters were determined \cite{ckmt} from a joint fit of the 
$\sigma^{tot}_{\gamma p}$ data and the NMC data \cite{nmce665} on
the proton structure function in the region 
$1 GeV^2 \leq Q^2 \leq 5 GeV^2$, and a very good
description of the experimantal data available was obtained. 

\section{Structure functions at high $Q^2$}
The next step in this approach is to introduce the QCD evolution in
the partonic distributions of the CKMT model and thus to 
determine the structure
functions at higher values of $Q^2$. For this, the
evolution equation in two loops in the $\overline{\mbox{MS}}$ 
scheme with 
$\Lambda=200 MeV$ was used \cite{ckmt}.

As starting point for the QCD evolution, the value $Q^2_0=2 GeV^2$,
where the logarithmic derivative in $Q^2$ of $F_2(x,Q^2)$ in 
the model is
very close to that obtained from the QCD evolution equation, was chosen.

The results obtained by taking into account the QCD evolution in
this way are \cite{ckmt} in a very good agreement with 
the experimental data on
$F_2(x,Q^2)$ at high values of $Q^2$.

Although the fit of the NMC data was restricted to the region 
$1 GeV^2 \leq Q^2 \leq 5 GeV^2$, it is interesting to mention that a
good fit can also be obtained with the model up to 
$Q^2\sim 10 GeV^2$. This allows to start the perturbative QCD
evolution at larger values of $Q^2$.

\section{Description of the HERA data on $F_2$ at low $Q^2$}
When the CKMT model was first used~\cite{ckmt} to fit 
the available 
experimental data on
$F_2$, the lack of experimental data at low and
moderate $Q^2$ limited the accuracy in the determination 
of the values of the
parameters in the model. Later on, the publication of the 
new experimental
data \cite{h1,zeus} on $F_2$ from HERA at low and
moderate $Q^2$ provided the opportunity to include in the fit of 
the parameters
of the model experimental points from the kinematical
region where the parametrization should give a good description 
without need of
any perturbative QCD evolution.

\begin{figure}[ht]
\begin{center}
\centering
\vskip -0.5cm
\hskip 4 cm
\epsfig{figure=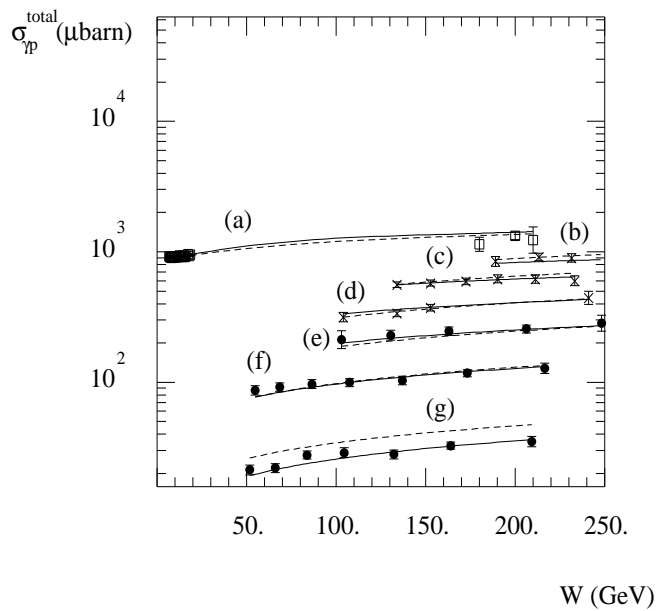,
bbllx=0,bblly=0,
bburx=504,bbury=720,
height=15.cm,width=10.cm}
\end{center}
\vskip -5.5cm
\caption{$\sigma^{total}_{\gamma p}$ and
$\sigma^{total}_{\gamma^* p}$
in $\mu$barns vs W in GeV for different values of $Q^2$.
 Theoretical fits
have been
obtained with the CKMT
model (full line) and the modified version of the CKMT m
odel (dashed line).
Points at (a), $Q^2$=0.$GeV^2$ (*8.), (b), $Q^2$=0.15$Ge
V^2$ (*6.),
(c), $Q^2$=0.25$GeV^2$ (*5.), (d), $Q^2$=0.5$GeV^2$ (*4.
),
(e), $Q^2$=0.8$GeV^2$ (*3.), (f), $Q^2$=1.5$GeV^2$ (*2.)
,
and (g), $Q^2$=3.5$GeV^2$ (*1.).
Experimental data on $F_2$ ($\sigma^{total}_{\gamma^* p}
$) are from
references (3), (4), and  experimental data on
$\sigma^{total}_{\gamma p}$ are
from references~(7).
\label{fig:fig1}}
\end{figure}

\begin{figure}[ht]
\begin{center}
\centering
\vskip -1.cm
\hskip 4 cm
\epsfig{figure=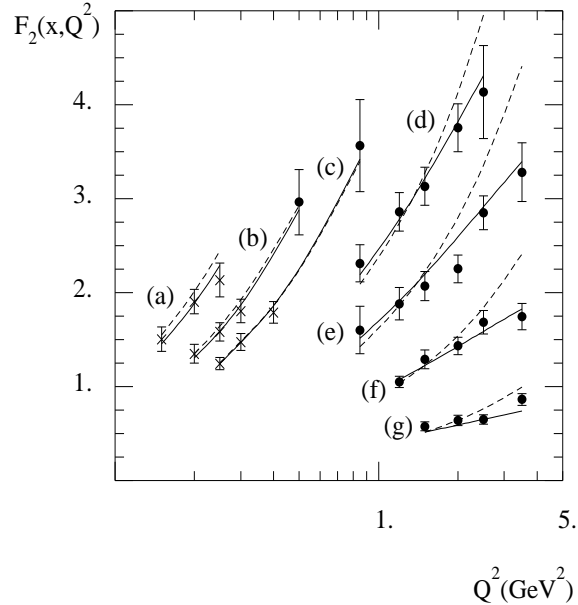,
bbllx=0,bblly=0,
bburx=504,bbury=720,
height=15.cm,width=10.cm}
\end{center}
\vskip -6cm
\caption{$F_2$($x$,$Q^2$) vs $Q^2$ (in $GeV^2$)
 for different values
of x. Theoretical fits have been
obtained with the CKMT
model (full line) and the modified version of the CKMT
model (dashed line).
Experimental points at (a), from left to right,
x=0.42$\cdot$$10^{-5}$,
x=0.44$\cdot$$10^{-5}$,
and x=0.46$\cdot$$10^{-5}$~(*8.);
(b), from left to right,  x=0.85$\cdot$$10^{-5}$,
x=0.84$\cdot$$10^{-5}$,
x=0.83$\cdot$$10^{-5}$, and x=0.86$\cdot$$10^{-5}$~(*6.)
;
(c), from left to right, x=0.13$\cdot$$10^{-4}$, and thr
ee
points at
x=0.14$\cdot$$10^{-4}$~(*5.); (d),
x=0.5$\cdot$$10^{-4}$~(*4.); (e),
x=0.8$\cdot$$10^{-4}$~(*3.);
(f), x=0.2$\cdot$$10^{-3}$~(*2.); (g),
x=0.5$\cdot$$10^{-3}$~(*1.);
(h) x=0.52$\cdot$$10^{-2}$~(*1.);
(i) x=0.13$\cdot$$10^{-1}$~(*1.).
Experimental points for $F_2$ are from
references~(3),~(black circles),~(4),~(crosses),
and~(6),~(black squares
for the E665 points, and black diamonds for the
NMC points).
\label{fig:fig2}}
\end{figure}

Thus, one proceeded \cite{newpaper} as one had done in the 
previous fit \cite{ckmt}, but by adding
the above mentioned
experimental data on $F_2$ from H1 and ZEUS at low and moderate $Q^2$, to
those \cite{nmce665} from NMC and E665 collaborations, and to data \cite{cross}
on
cross-sections for real photoproduction, into a global fit which 
allowed 
the test
of the model in wider regions of $x$ and $Q^2$. One took as 
initial condition
for the values of the different parameters
those obtained in the previous fit \cite{ckmt}.
The result of the new common fit to $\sigma^{tot}_{\gamma p}$
and $F_2$ is presented in figures \ref{fig:fig1} 
and \ref{fig:fig2}, and the final values of the 
parameters
can be found in Table\ref{tab:tab1}(b). The
quality of the
description of all the experimental data provided by the 
CKMT model, and, in
particular, of the
the new experimental data from HERA is very good, with a value
of $\chi^2/d.o.f.$ for the global fit, $\chi^2/d.o.f.$=106.95/167, 
where
the statistical and
systematic
errors have been treated in quadrature, and where the relative normalization
among all the experimental data sets has been taken equal to 1.
\noindent
Moreover, since the small-$x$ HERA experiments allowed for 
the first time the
experimental study of the question of the interplay between soft and hard
physics,
the model, which basically has only power dependence
on $Q^2$, was modified to include a logarithmic
dependence on
$Q^2$ as the one predicted
asymptotically by perturbative QCD.

To include the logarithmic dependence on $Q^2$ in our 
model, we take into
account that the behavior of $F_2$ at small-$x$ 
in QCD is given by the
singularities of the moments of the structure 
functions \cite{rujula}, the 
rightmost singularity giving the leading behavior. 
Thus,
the following factors \cite{rujula} \cite{paco}, 
which correspond
to the moments of the structure functions
in the language of the OPE expansion, and
can be calculated by
the convolution in rapidity of the hard-upper part with
the soft-lower
part of the leptoproduction diagram, were introduced in 
the expression that the CKMT model gives for $F_2$:
\begin{equation}
\left({\alpha_s(Q^2_0)\over\alpha_s(Q^2)}\right)^{d_i
(n_i)}, i=S,NS,
\label{eq:factors}
\end{equation}
\noindent
where the strong coupling constant is taken as
\begin{equation}
\alpha_s(Q^2) = {4\pi\over\beta_0\cdot log\left({Q^2+
M^2\over
\Lambda_{QCD}^2}\right)},
\label{eq:factors1}
\end{equation}
\noindent
with M$\sim$1GeV, a hadronic mass included 
in \ref{eq:factors1}
to avoid the
singularity in
$\alpha_s$ when $Q^2$$\rightarrow \Lambda_{QCD}^2$,
$\Lambda_{QCD}$=0.2 GeV, and
$\beta_0$=11$-{2\over3}n_f$ (in the calculations
a number of flavors
$n_f$=3 was used). The exponents $d_S(n_S)$ 
and $d_{NS}(n_{NS})$
in~\ref{eq:factors} are
proportional to
the largest eigenvalue of the 
anomalous dimension matrix,
and to the anomalous
dimension, respectively:
\begin{equation}
d_S(n_S) \sim {d_0\over4(n_S-1)}-d_1,
\label{eq:factors2}
\end{equation}
\noindent
with
\begin{equation}
d_0 = {48\over\beta_0}, d_1 = {11+{2\over 27}n_f\over
\beta_0},
\end{equation}
\noindent
and
\begin{equation}
d_{NS}(n_{NS}) = {16\over33-2n_f}\cdot\left({1\over2n
_{NS}(n_{NS}+1)}
+{3\over4}-S_1(n_{NS})\right),
\label{eq:factors3}
\end{equation}
\noindent
with
\begin{equation}
S_1(n_{NS}) = n_{NS}\cdot\sum^{\infty}_{k=1}
{1\over k(k+n_{NS})}.
\end{equation}
\par

\noindent
Thus, 
we modify \ref{eq:eq1}
in the following way \cite{paco}
\begin{equation}
F_2(x,Q^2) = \left({\alpha_s(Q^2_0)\over\alpha_s(Q^2)}\right)^{d_S(n_S)}
\cdot F_S(x,Q^2) +
\left({\alpha_s(Q^2_0)\over\alpha_s(Q^2)}\right)^{d_{NS}(n_{NS})}
\cdot F_{NS}(x,Q^2).
\label{eq:eq7}
\end{equation}
The exponents of the new factors in \ref{eq:eq7}, 
$d_S(n_S)$ and $d_{NS}(n_{NS})$,
give us the singularities in $n_i$, $i$=$S$,$NS$, of the momenta, which, as we
mentioned
above, control the QCD small-$x$ behavior of $F_2$. Therefore, 
in the CKMT model,
these exponents
have to be evaluated (see \ref{eq:eq2}  and \ref{eq:eq4}), at
$n_S$=1+$\Delta(Q^2$$\rightarrow$$\infty$)=
1+$\Delta_0$(1+$\Delta_1$), and at $n_{NS}$=$\alpha_R$, respectively.

Then, this modified version of the CKMT 
parametrization of $F_2$ was used to 
repeat the fit of the same experimental data, including the HERA data on $F_2$ at
small and moderate $Q^2$. As starting point for
the QCD evolution, one takes the same value that
$Q^2_0$=2.$GeV^2$ was used to fix the
normalization of the valence component.
The result of this second fit is also presented in
figures~\ref{fig:fig1} and~\ref{fig:fig2}, and the final values 
of the parameters in 
the model 
are given
in Table\ref{tab:tab1}(c). 

As it can
be seen in the figures~\ref{fig:fig1} and~\ref{fig:fig2}, the 
quality of this second fit is 
also reasonable,
although the value of $\chi^2/d.o.f.$ ($\chi^2/d.o.f.$=453.19/167), 
is now
appreciably higher than in the fit obtained
with the non-modified version of the CKMT model.
\begin{table}[ht]
\caption{Values of the parameters in the CKMT model obtained in
former fits, (a), in the fit in which also the low $Q^2$ HERA data
have been included, (b), and in the fit to the same data obtained with the
modified version of the CKMT model in which a logarithmic dependence of $F_2$
on $Q^2$ has been taken into account, (c). All dimensional parameters are given
in $GeV^2$. The
valence counting rules provide the following values of $B_u$ and $B_d$, for the
proton case, when fixing their normalization at $Q_0^2$=2.$GeV^2$: (a)
$B_u$=1.2064, $B_d$=0.1798; (b) $B_u$=1.1555, $B_d$=0.1722; (c) $B_u$=0.6862,
$B_d$=0.09742. In previous fits, (a), the
parameter $\Delta_1$ had been fixed to a value $\Delta_1$=2.\label{tab:tab1}} 
\vspace{0.2cm}
\centering
\vskip 1.cm
\hskip 1.cm
\footnotesize
\vbox {\offinterlineskip
\hrule
\halign{&\vrule#&
\strut\quad\hfil#\quad\cr
height2pt&\omit&&\omit&&\omit&&\omit&\cr
&CKMT model\hfil&&{\bf (a)\hfil}&&{\bf (b)\hfil}&&{\bf (c)\hfil}&\cr
height2pt&\omit&&\omit&&\omit&&\omit&\cr
\noalign{\hrule}
height2pt&\omit&&\omit&&\omit&&\omit&\cr
&A\hfil&&0.1502\hfil&&0.1301\hfil&&0.1188\hfil&\cr
height2pt&\omit&&\omit&&\omit&&\omit&\cr
\noalign{\hrule}
height2pt&\omit&&\omit&&\omit&&\omit&\cr
&a\hfil&&0.2631\hfil&&0.2628\hfil&&0.07939\hfil&\cr
height2pt&\omit&&\omit&&\omit&&\omit&\cr
\noalign{\hrule}
height2pt&\omit&&\omit&&\omit&&\omit&\cr
&$\Delta_0$\hfil&&0.07684\hfil&&0.09663\hfil&&0.1019\hfil&\cr
height2pt&\omit&&\omit&&\omit&&\omit&\cr
\noalign{\hrule}
height2pt&\omit&&\omit&&\omit&&\omit&\cr
&$\Delta_1$\hfil&&2.0 (fixed)\hfil&&1.9533\hfil&&1.2527\hfil&\cr
height2pt&\omit&&\omit&&\omit&&\omit&\cr
\noalign{\hrule}
height2pt&\omit&&\omit&&\omit&&\omit&\cr
&$\Delta_2$\hfil&&1.1170\hfil&&1.1606\hfil&&0.1258\hfil&\cr
height2pt&\omit&&\omit&&\omit&&\omit&\cr
\noalign{\hrule}
height2pt&\omit&&\omit&&\omit&&\omit&\cr
&c\hfil&&3.5489\hfil&&3.5489 (fixed)\hfil&&3.5489 (fixed)\hfil&\cr
height2pt&\omit&&\omit&&\omit&&\omit&\cr
\noalign{\hrule}
height2pt&\omit&&\omit&&\omit&&\omit&\cr
&b\hfil&&0.6452\hfil&&0.3840\hfil&&0.3194\hfil&\cr
height2pt&\omit&&\omit&&\omit&&\omit&\cr
\noalign{\hrule}
height2pt&\omit&&\omit&&\omit&&\omit&\cr
&$\alpha_R$\hfil&&0.4150\hfil&&0.4150 (fixed)\hfil&&0.5872\hfil&\cr
height2pt&\omit&&\omit&&\omit&&\omit&\cr}
\hrule}
\end{table}

\section{Description of the Caldwell-plot}
The so-called Caldwell-plot shows~\cite{caldw} the logarithmic slope 
of the structure
function $F_2$, $dF_2/dlnQ^2$, derived from the ZEUS data, as a function of
$x$, by fitting
$F_2\sim a+blnQ^2$ in bins of fixed $x$, using only statistical errors. This
plot allows the study of the QCD scaling violations of $F_2$, and, in 
particular, in the small-$x$ domain now accessible at HERA, where $dF_2/dlnQ^2$
is directly related to the gluon density, 
can be an useful tool to
investigate down to which value of $Q^2$ the perturbative NLO DGLAP QCD
predictions give a good description of the $F_2$ data. Thus, the logarithmic
slope $dF_2/dlnQ^2$ can be used to investigate the fundamental question of
the interplay between the soft and the hard physics.

As it can be seen in reference~(10), for values of 
$x$ down to $3\cdot 10^{-4}$,
the slopes are increasing as $x$ decreases, but at lower values of $x$ and $Q^2$
the slopes decrease, what seems to indicate a deviation from 
the perturbative
behavior of the hard regime.

Also in the reference \cite{caldw} one can see how both Regge 
based parametrizations with a
constant effective value for the Pomeron intercept, like the 
Donnachie-Landshoff Regge fit or the ZEUSREGGE fit, and 
pure perturbative NLO
QCD predictions, like GRV94 NLO QCD fit or the ZEUSQCD fit, 
fail in describing correctly the
experimental data in the whole kinematical region. While 
the Donnachie-Landshoff
and the ZEUSREGGE fits do not describe the data for 
values of $x$ larger than
$\sim 10^{-5}$, GRV and ZEUSQCD do not follow the data 
when one goes to 
values of $x$ 
smaller than $\sim 6\cdot 10^{-5}$ 
(see~(10) and references therein for more details on this
discussion).

On the other side, the CKMT model described above, based 
on the Regge behavior,
but with a $Q^2$-dependent Pomeron 
intercept, describes \cite{prog} the data 
in the region
of low $Q^2$, and when taken as the initial condition for 
the QCD
evolution equations, provides a complete 
description of the experimental
results in the whole ranges of $x$ and $Q^2$.

As a matter of fact, by using the formulae 
of the pure CKMT model in section 1, 
when $x$ is kept fixed one can 
write \cite{prog} for 
the CKMT model the 
slope $dF_2/dlnQ^2$ as:
\begin{equation}
\begin{array}{rcl}
\frac{dF_2(x,Q^2)}{dlnQ^2} & = & F_S(x,Q^2)
[\frac{\Delta_2}{Q^2+\Delta_2}
\left(\Delta(Q^2)-\Delta_0\right)
ln\frac{Q^2}{x(Q^2+a)}\\
&&
+\frac{c}{Q^2+c}
\left(n(Q^2)-\frac{3}{2}\right)ln(1-x)
+\frac{a\left(1+\Delta(Q^2)\right)}{Q^2+a}]\\
&+& F_{NS}(x,Q^2)
[\frac{c}{Q^2+c}\left(n(Q^2)-\frac{3}{2}\right)ln(1-x)\\
&&
+\frac{b\alpha_R(0)}{Q^2+b}],\\ & &
\end{array}
\label{eq:eq8}
\end{equation}
that in the limit $Q^2\rightarrow 0$ takes the form
\begin{equation}
\begin{array}{rcl}
\frac{dF_2(x,Q^2)}{dlnQ^2} & \sim & 
\left(1+\Delta_0\right) F_S(x,Q^2)\\
&+& \alpha_R(0) F_{NS}(x,Q^2).\\
&&
\end{array}
\label{eq:eq9}
\end{equation}
Also, if one considers the case when W is fixed one can 
take $x\sim cte\cdot Q^2$, 
and then,
up to constant factors, one gets:
\begin{equation}
\begin{array}{rcl}
\frac{dF_2(x,Q^2)}{dlnQ^2} & = &
F_S(x,Q^2)
[-\frac{\Delta_2}{Q^2+\Delta_2}\left(\Delta(Q^2)-\Delta_0\right)
ln(Q^2+a)\\
&&
-\Delta(Q^2)+\frac{c}{Q^2+c}\left(n(Q^2)-\frac{3}{2}\right)ln(1-Q^2)\\
&&
-\frac{Q^2n(Q^2)}{1-Q^2}+\frac{a(1+\Delta(Q^2))}{Q^2+a}]\\
&+& F_{NS}(x,Q^2)
[\frac{c}{Q^2+c}(n(Q^2)-\frac{3}{2})ln(1-Q^2)\\
&&
+\frac{b\alpha_R(0)}{Q^2+b}+(1-\alpha_R(0))-\frac{Q^2n(Q^2)}{1-Q^2}].\\
&&
\end{array}
\label{eq:eq10}
\end{equation}
Now, if one takes W fixed with $Q^2\sim x\rightarrow 0$, 
one can easily see 
that this equation simply reduces to:
\begin{equation}
\frac{dF_2(x,Q^2)}{dlnQ^2} \sim F_2(x,Q^2).
\label{eq:eq11}
\end{equation}
One has to note that both equations \ref{eq:eq9} 
and \ref{eq:eq11} are
valid for any well-behaved parametrization of $F_2$ 
(i.e., any parametrization fulfilling the relation 
in equation \ref{eq:eq6}).

Taking into account the general features of the CKMT model
described above, we use \cite{prog} the CKMT model  
to describe 
the experimental
data in the region of low $Q^2$ 
($0< Q^2< Q^2_0=2.GeV^2$), 
and
then we take this parametrization as the initial condition at 
$Q^2_0=2.GeV^2$, to be
used in the QCD evolution equation to obtain a description of the
experiment at values of $Q^2$ higher than $Q^2_0=2.GeV^2$. We
present our results in the shape of both the $dF_2/dlnQ^2$ and the
$dlnF2/dln(1/x)$ slopes in order to compare with the experimental
data when presented in the so-called Caldwell-plot.

The way we proceed to calculate $F_2$, and the derivatives 
$dF_2/dlnQ^2$ and $dlnF2/dln(1/x)$ is the following (see
reference~(12) and the appendix there for all the technical details 
on how the QCD evolution has been performed):
\begin{itemize}

\item  In the region $0< Q^2\le Q^2_0=2.GeV^2$ we use the pure
CKMT model for $F_2$.

\item  For $Q^2_0< Q^2\le charm$ $threshold$ \cite{grv}, 
we make the QCD
evolution of $F_2$ at NLO in the $\overline{\mbox{MS}}$ 
scheme for a 
number of flavours $n_f=3$, and we take 
as the starting parametrization the CKMT
one at $Q^2_0=2.GeV^2$. 

\item  When $charm$ 
$threshold < Q^2\le \bar{Q}^2=50. GeV^2$, also
the QCD evolution of $F_2$ is implemented at NLO in the 
$\overline{\mbox{MS}}$
scheme for a
number of flavours $n_f=3$, using the parton distribution
functions for the $u,d,s$ quarks, and by including the charm
contribution via photon-gluon fusion.

\item  For values of $Q^2> \bar{Q}^2$, QCD evolution is computed
at NLO in the $\overline{\mbox{MS}}$
scheme, but now with a
number of flavours $n_f=4$, and by using the parton distribution
functions for the $u,d,s$, and $c$ quarks.

\end{itemize}

One has to note that in the treatment of the charm contribution we
have followed reference (13).

\begin{figure}[ht]
\begin{center}
\centering
\hskip -3cm
\epsfig{file=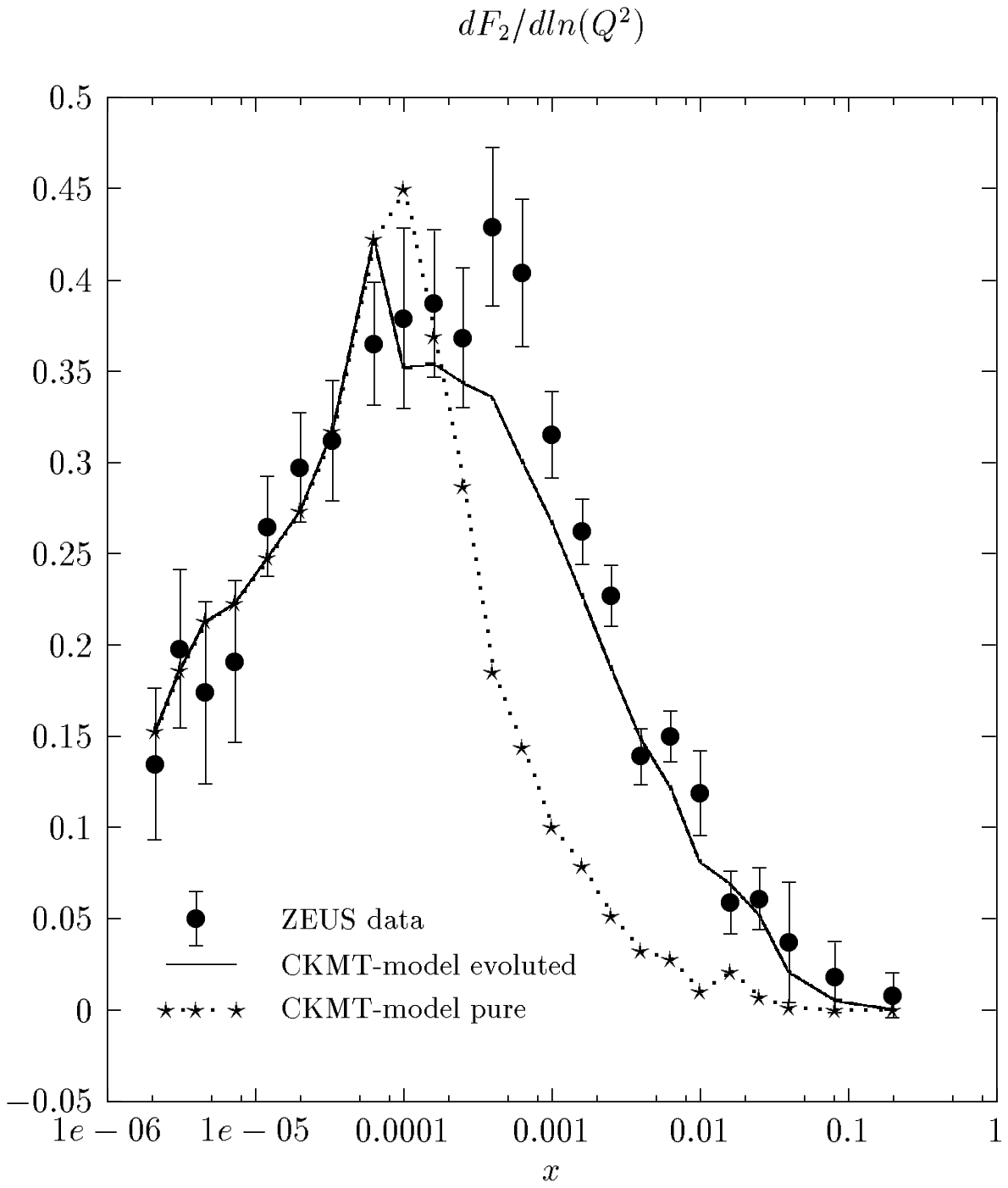,bbllx=0,bblly=0,
bburx=504,bbury=720,height=15.cm,width=10.cm}
\end{center}
\vskip -6.5cm
\caption{$dF_2/dlnQ^2$ as a function of $x$ computed
in the CKMT model (see reference (12) for details 
on the calculation), compared with the fit of the
ZEUS $F_2$
data in bins of $x$ to the form $a+blnQ^2$
(see reference (10) and references therein for
more details on the data and the experimental fit).
\label{fig:fig3}}
\end{figure}

\begin{figure}[ht]
\begin{center}
\centering
\hskip -3cm
\epsfig{file=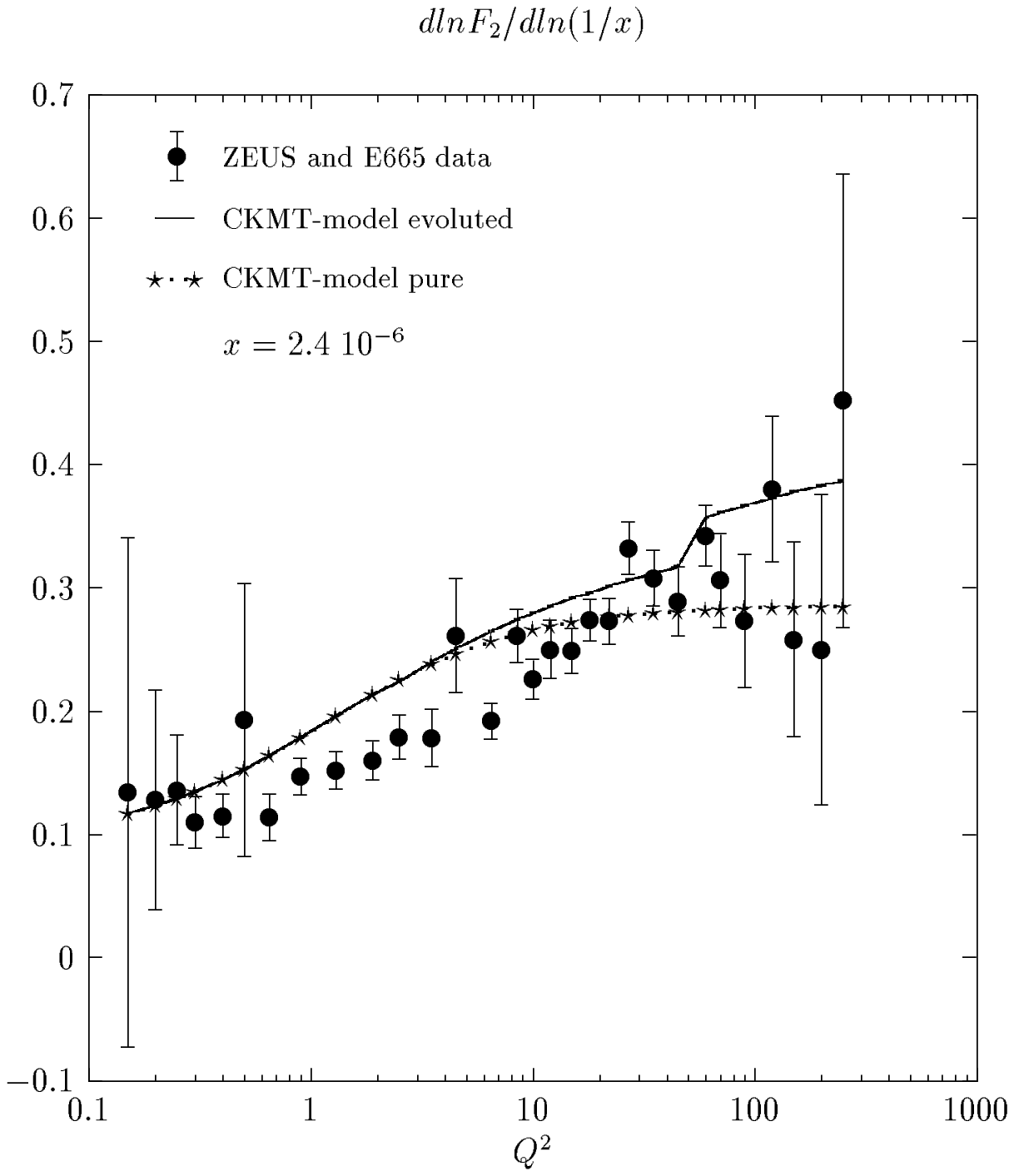,bbllx=0,bblly=0,
bburx=504,bbury=720,height=15cm,width=10.cm}
\end{center}
\vskip -6.5cm
\caption{$dlnF_2/dln(1/x)$ as a function of $Q^2$ calculated
in the CKMT model, and compared to the fit 
$F_2=Ax^{-\lambda_{eff}}$ of the ZEUS (10) and E665 (6) 
data with $x<0.01$
For details on the CKMT calculation, see reference (12).
\label{fig:fig4}}
\end{figure}

\begin{figure}[ht]
\begin{center}
\centering
\hskip -3cm
\epsfig{file=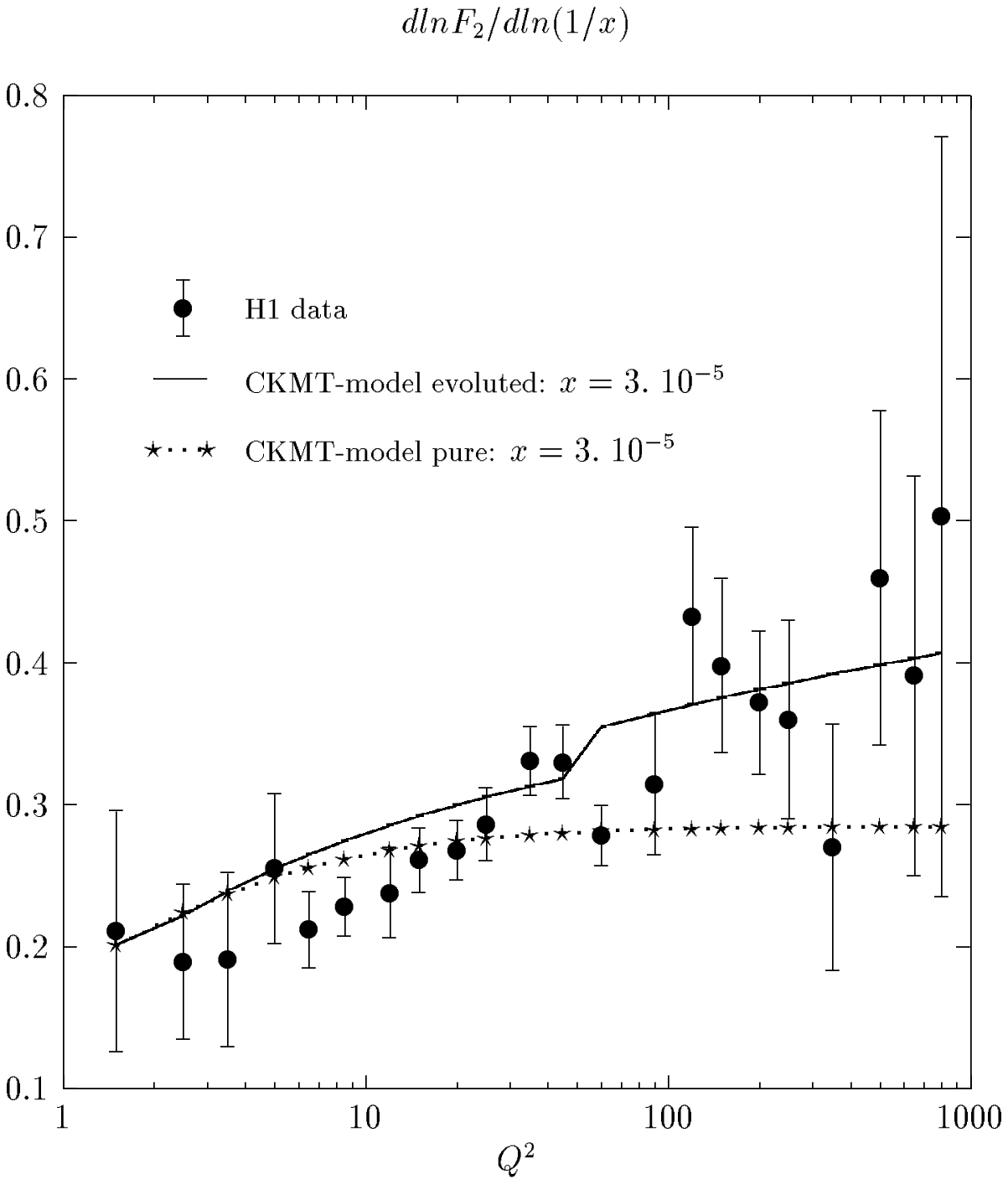,bbllx=0,bblly=0,
bburx=504,bbury=720,height=15.cm,width=10.cm}
\end{center}
\vskip -6.5cm
\caption{$dlnF_2/dln(1/x)$ as a function of $Q^2$ calculated
in the CKMT model, and compared to the fit 
$F_2=Ax^{-\lambda_{eff}}$ of the H1 
data (11).
For details on the CKMT calculation, see reference (12).
\label{fig:fig5}}
\end{figure}

\begin{figure}[ht]
\begin{center}
\vskip .5cm
\centering
\hskip -4cm
\epsfig{file=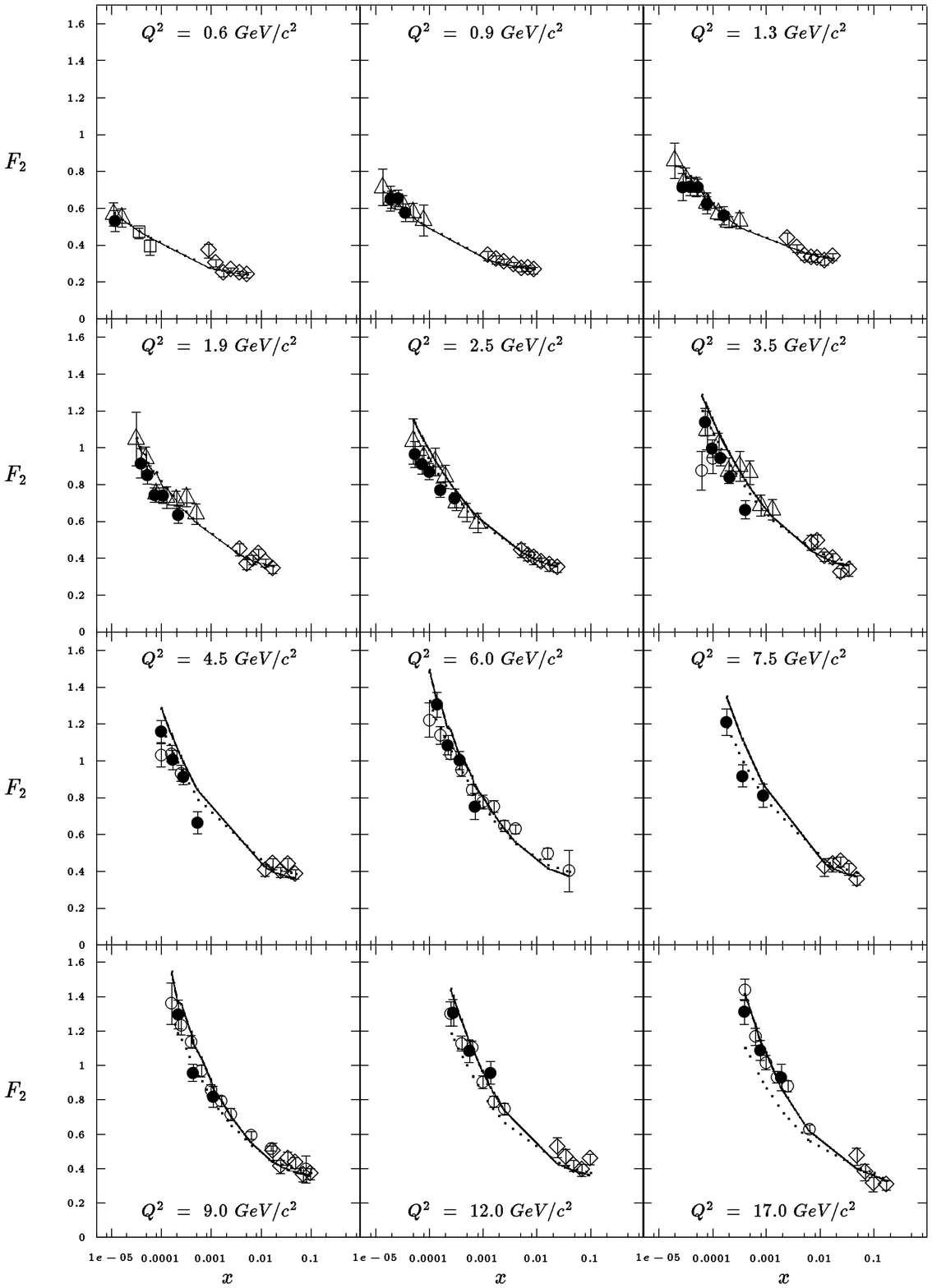,bbllx=0,bblly=0,
bburx=504,bbury=720,height=16cm,width=14cm}
\end{center}
\vskip -2cm
\caption{$F_2$ as a function of $x$ computed in the CKMT model (12) 
for twelve different values of $Q^2$, and compared with the following
experimental data (see (10) for the experimental references): 
ZEUS SVX95 (black circles), H1 SVX95 (white
triangles), ZEUS BPC95 (white squares), E665 (white diamonds), and ZEUS
94 (white circles). The dotted line is the theoretical result obtained
with the pure CKMT model, and the solid line is the result obtained with
the QCD-evoluted CKMT model.
\label{fig:fig6}}
\end{figure}

The results we have obtained are presented in figures 
\ref{fig:fig3} to \ref{fig:fig6}. In
figure \ref{fig:fig3} (Caldwell-plot), the slope 
$dF_2/dlnQ^2$ is shown as a function of $x$, and compared with
the $a+blnQ^2$ fit to the ZEUS $F_2$ data in bins of $x$. 

Figures \ref{fig:fig4} and \ref{fig:fig5} show the slope
$dlnF2/dln(1/x)$ as a function of $Q^2$ compared to the fits
$F_2=Ax^{-\lambda_{eff}}$ of the the ZEUS and H1 data, 
respectively. In Figure \ref{fig:fig4}, as the $x$ range of the BPC95
data is restricted, also the E665 \cite{nmce665} data 
were included in (10), and are now also taken into account. 
The interest of these figures is clear, since
this slope can be interpreted as 
the effective $\lambda$ of the
Pomeron exchange, $\lambda_{eff}=dlnF2/dln(1/x)$. 
In the experimental fits, each $Q^2$ bin corresponds to
a average value of $x$, $<x>$, calculated from the mean value of
$ln(1/x)$ weighted by the statistical errors of the corresponding
$F_2$ values in that bin. Even though we can proceed as in the
experimental fits, and we get a very good agreement with the data, 
since the estimation of $<x>$ is in some
sense artificial and arbitrary, and introduces unphysical wiggles
when drawing one full line connecting the different bins, 
we prefered
to make for all the $Q^2$ bins in this figures the choice of the 
smallest $x$ in the data 
instead of considering a different $<x>$ 
for each $Q^2$. This choice is based
on the fact that the ansatz $\lambda_{eff}=dlnF2/dln(1/x)$ is
actually valid for small $x$, 
and results in a smooth curve except for the
jump in the region around $Q^2\sim 50 GeV^2$, where the evolution
procedure changes (again, see reference (12) for more details). 

Finally, figure \ref{fig:fig6} is the compilation of the
behavior of $F_2$ as a function of $x$ for twelve different
values of $Q^2$ (from $Q^2=0.6 GeV^2$ to $Q^2=17. GeV^2$), 
corresponding to the values presented by the ZEUS Collaboration in
reference (10). 

A very good agreement with the experiment is obtained for
all the $x$ and $Q^2$ values, showing that the experimental data can be
described by using as initial condition for the QCD evolution equation
a model of $F_2$ where the shadowing effects which are important at low
values of $Q^2$ are included.

\section{Conclusions}
The CKMT model for the parametrization of 
the nucleon
structure functions
provides a very good description of all the
available experimental data on $F_2(x,Q^2)$ at low and moderate $Q^2$,
including the more
recent small-$x$ HERA points. 
A second fit to the same data obtained with a
modified version of the model in
which a
logarithmic
dependence on $Q^2$ is included, has been 
also presented. Even though the quality of
this second description is reasonable, its $\chi^2/d.o.f.$ is appreciably
higher
than that corresponding to the fit obtained with the non-modified version of
the CKMT model.

Finally, the CKMT model for $F_2$ has been used as the initial 
condition in the QCD-evolution equation to
describe the HERA experimental data presented 
in the so-called Caldwell-plot,
where the $x$-dependence of the logarithmic slope of the 
structure function, $dF_2/dlnQ^2$, is shown 
for different $Q^2$ bins, and in the plot of the 
$Q^2$-dependence of the $\lambda_{eff}$, i.e., of the $Q^2$ behavior of
the slope $dlnF_2/dln(1/x)$, now for different bins of $x$. 
The obtained results show that the
available experimental data can be described by performing the QCD
evolution of a model of $F_2$ where the shadowing effects which are 
important at low values of $Q^2$ are included. 

\section*{Acknowledgments}
C. Merino wants to thank Sandra Oliveira for her help before and 
during the conference, and
the members 
of the Local Organizing Committee 
for the perfect organization and the nice atmosphere of this school.

\end{document}